\documentclass{article}
\usepackage[utf8]{inputenc}
\usepackage{graphicx}

\makeatletter

\@ifundefined{date}{}{\date{}}
\usepackage{authblk}
\usepackage{setspace}
\usepackage{amsmath}

\title{Neural-network-powered pulse reconstruction from one-dimensional interferometric cross-correlation traces}
\author[1,2]{Pavel V. Kolesnichenko}
\author[1,2]{Donatas Zigmantas}
\affil[1]{\textit{Chemical Physics, Lund University, P.O. Box 124, 221 00 Lund, Sweden}}
\affil[2]{\textit{NanoLund, Lund University, P.O. Box 118, 221 00 Lund, Sweden}}

\makeatother

\usepackage[sorting=none]{biblatex}
\begin{document}
\maketitle \doublespacing 

{\centering\subsection*{Abstract}}

 Any ultrafast optical spectroscopy experiment is usually accompanied by the necessary routine of ultrashort-pulse characterisation. The majority of pulse characterisation approaches solve either a one-dimensional (e.g. via interferometry) or a two-dimensional (e.g. via frequency-resolved measurements) problem. 
Solution of the two-dimensional pulse-retrieval problem is generally more consistent due to problem's over-determined nature. In contrast, the one-dimensional pulse-retrieval problem is impossible to solve unambiguously as ultimately imposed by the fundamental theorem of algebra. In cases where additional constraints are involved, the one-dimensional problem may be possible to solve, however, existing iterative algorithms lack generality, and often stagnate for complicated pulse shapes
. Here we use a deep neural network to unambiguously solve a constrained one-dimensional pulse-retrieval problem and show the potential of fast, reliable and complete pulse characterisation using interferometric cross-correlation time traces (determined by the pulses with partial spectral overlap).

\section{Introduction}

Optical pulse characterisation is a necessary step of all ultrafast spectroscopy experiments performed prior to any measurement of dynamics in media of interest. 
The most precise and reliable pulse characterisation techniques are considered those that allow for pulse retrieval from two-dimensional (2D) data-sets acquired, for example, by frequency-resolved optical gating (FROG) \cite{Kane1993}, multiphoton intrapulse interference phase scan (MIIPS) \cite{Lozovoy2004}, or dispersion scan (D-Scan) \cite{Miranda2011}. This reliability stems from the over-determination of the corresponding pulse-retrieval problems: a much greater number of points (of the order of $\sim N^2$) representing a spectrally- and time-resolved 2D data-set is mapped onto a much lesser number of only $\sim N$ points describing pulse electric field as a one-dimensional (1D) function of time.

It is still, nevertheless, useful to be able to reliably reconstruct electric fields of ultrashort pulses from 1D data-sets due to faster data-acquisition times and potentially simpler experimental settings. Due to the fundamental theorem of algebra, however, this is impossible to achieve, and additional information is required \cite{Kane1993,Trebino2000}.

Amongst 1D pulse retrieval approaches, spectral phase interferometry for direct electric-field reconstruction (SPIDER) \cite{Iaconis1998} is arguably the most successful technique capable of unambiguous pulse retrieval. 
In this approach, the additional information needed for solving the 1D pulse-retrieval problem is encoded by using a significantly chirped reference pulse for sum-frequency-generation signal (between the chirped pulse and the two replicas of the pulse to be characterised). The experimental implementation of the SPIDER method, however, is complex, involves two delay stages and requires highly precise alignment. Other methods based on measuring triple intensity correlations \cite{Kakarala1993} also uniquely yield pulse shapes, but also at an expense of increased complexity.

Acquiring 1D time-domain traces of intensity correlations between two pulses only, on the other hand, is very easy to implement experimentally (as only one delay stage is required), and numerous attempts to reconstruct pulse shapes from such traces were previously made \cite{Diels1985,Diels1985a,DIELS:87,Naganuma1989,Naganuma1989a,Naganuma1990,Yan1991,Peatross1998}. Peatross \textit{et al.} \cite{Peatross1998} were able to reconstruct electric fields of ultrashort pulses from intensity autocorrelation traces using the Gerchberg-Saxton algorithm \cite{Gerchberg72}. They note, however, that their approach suffers from non-trivial ambiguities as more than one pair of electric field amplitude and phase can lead to essentially same autocorrelation trace. Moreover, their algorithm does not outperform the iterative FROG algorithm \cite{Kane1993} or recently reported common pulse retrieval algorithm (COPRA) \cite{Geib2019}, and can stagnate. Later, Chung and Weiner \cite{Chung2001}, numerically proved that significantly different pulse shapes can lead to exactly the same intensity autocorrelation.

It was also shown previously that in cases where different pulse shapes lead to same intensity autocorrelation traces, interferometric autocorrelation traces are similar but not equivalent \cite{Chung2001}. Thus, interferometric correlations are also sensitive to phase of electric field, which inspired attempts to find sufficient number of constraints to solve the corresponding pulse-retrieval problem. Diels \textit{et al.} \cite{Diels1985,Diels1985a} suggested that laser-pulse spectrum together with second-order intensity and interferometric autocorrelation traces (acquired, for example, by mixing two pulses in a nonlinear crystal via second-harmonic generation) are sufficient to reconstruct pulse electric field, although no proof was given that their iterative algorithm yielded unique solutions. Later, a technique named as the Femto-nitpicker was introduced \cite{DIELS:87,Yan1991}. This technique was capable of retrieving pulse shapes from interferometric autocorrelation traces, which were supplemented by interferometric cross-correlation measurements obtained when a thick glass block was inserted into one of the interferometer's arms. However, the algorithm assumed only broadening effect of the glass block (and, in general, would fail for the case of pulse narrowing) and the accuracy of their iterative algorithm was lower for thinner glass blocks.

Naganuma \textit{et al.} suggested that for successful 1D pulse retrieval it is sufficient to measure intensity autocorrelation, second-harmonic interferogram, and field (first-order) interferogram \cite{Naganuma1989}. Experimentally, they have also supplemented interferometric autocorrelation measurements with those performed with a piece of glass inserted before the interferometer \cite{Naganuma1989a} or in one of its arms \cite{Naganuma1990}. The iterative algorithm tended, however, to stagnate, and their experimental approach did not find applications \cite{Trebino2000}. 

The listed attempts, nevertheless, suggest that such pulse characteristics as the amount of chirp, pulse amplitude time profile and frequency variation within the pulse all leave their fingerprints on interferometric correlation traces. It is, therefore, desirable to be able to use these fingerprints to reconstruct pulse shapes from measurements with the simplest experimental arrangement. Convolutional neural networks (CNNs) \cite{Fukushima1980,LeCun1999,Schmidhuber2015} were proven to be extremely powerful in recognizing fingerprints and patterns in large non-trivially complicated data-sets, and therefore they could also aid pulse retrieval from 1D interferometric correlation traces.

Previously, neural networks were shown to be successful in pulse retrieval from 2D data-sets: FROG traces \cite{Noble2020,Zahavy2018,Krumbuegel1996,Ziv2020}, D-Scan traces \cite{Kleinert2019}, speckles at the output of a multi-mode fiber \cite{Xiong2020}, and streak traces characterising attosecond pulses \cite{Zhu2020}. However, neural-network-assisted reconstruction of ultrashort pulses from 1D interferometric cross-correlation data-sets, to the best of our knowledge, have not been demonstrated previously, and our proof-of-principle work aims to fill this gap, as summarized below. 

Following developments described above, here we investigate, with the help of CNNs, the question of ambiguity of the retrieval of ultrashort pulse shapes from interferometric correlation traces
. We consider a more general case of interferometric \textit{cross}-correlation trace (with an unknown reference pulse) supplemented with spectra of both pulses and an interferometric cross-correlation trace obtained when the pulse to be diagnosed is affected by a glass plate. This supplemental data serves as the set of additional constraints to the 1D pulse retrieval problem. 
We synthesize a large data-set to train our neural network, and apply the trained model to the experimental data obtained using the dispersionless Michelson interferometer described previously \cite{Kolesnichenko:20}. We demonstrate that it is possible to unambiguously retrieve electric fields of ultrashort pulses from the four mentioned 1D data-sets. Importantly, no other knowledge than spectrum (which partially overlaps the spectrum of the pulse to be characterised) is required of the reference pulse; instead, we use the interferometric cross-correlation measurement (without the glass plate) itself as a reference measurement.

\section{Data preparation for neural-network training}

\begin{figure}[h!]
\centering \includegraphics[width=1\textwidth,height=0.8\textheight,keepaspectratio]{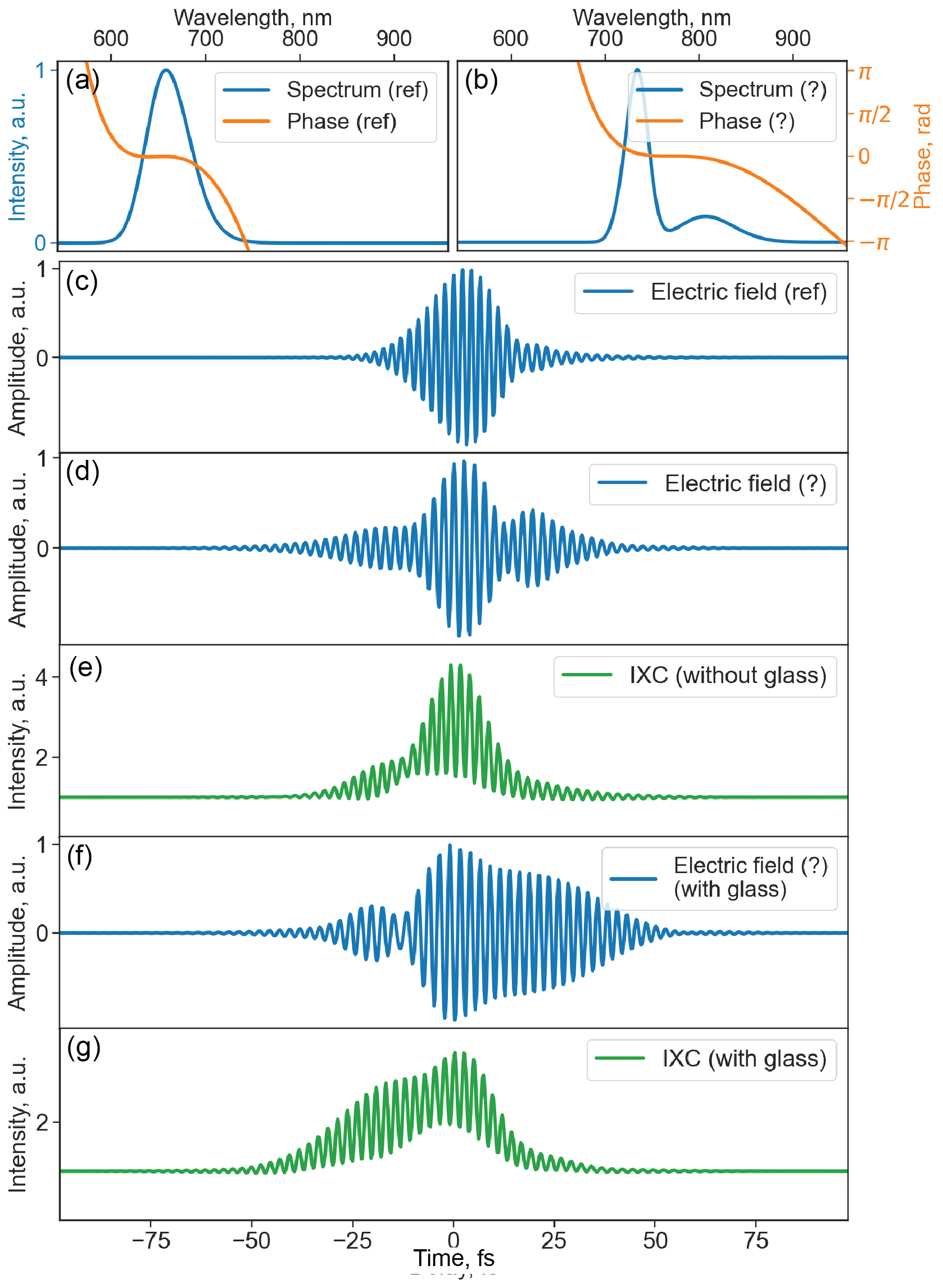}
\caption{Generated spectra and spectral phases of (a) the reference pulse and (b) the pulse to be characterised; time-domain electric fields of (c) the reference pulse, (d) the pulse to be characterised, and (e) the pulse to be characterised passed through a 2-mm fused silica glass plate; interferometric cross-correlation (IXC) traces with the 2-mm fused silica plate (f) out and (g) in.}
\label{fig:data_sample} 
\end{figure}

Numerical interferometric intensity cross-correlation traces ($I_{\textrm{IXC}}$) are generated via the expression
\begin{align}
I_{\textrm{IXC}}(t) = \int \Big|\big(E_{\textrm{ref}}(t) + E_?(t-\tau)\big)^m\Big|^2 \,d\tau,
\label{eqn:IC}
\end{align}
where $E_{\textrm{ref}}(t)$ is the electric field of the arbitrary reference pulse passed through the first arm of an interferometer; $E_?(t)$ is the electric field of the pulse to be characterised, passed through the second arm of the interferometer; $\tau$ is the delay between the two pulses; $m$ is the order of the cross-correlation signal, which we set to be equal to 2 (as second-order correlation measurements are most feasible to realise experimentally). The electric fields $E_{\textrm{ref}}(t)$ and $E_?(t)$ are obtained from randomly generated electric field amplitudes and phases defined in the frequency domain (see Supplementary material, Section S1, for more details). An example of a simulated interferometric correlation trace is given in Figure~\ref{fig:data_sample}: Figures~\ref{fig:data_sample}a,b show the spectra and the corresponding spectral phases of the two pulses used to simulate the correlation trace; the corresponding electric fields of the pulses are shown in Figure~\ref{fig:data_sample}c,d; the resultant cross-correlation trace obtained using the two pulses is shown in Figure~\ref{fig:data_sample}e.

In our work, we exploit sensitivity of machine-learning models to the presence of ambiguities in training data-sets \cite{10.5555/2998687.2998822,Gao2017}: a simple neural network, for example, will converge well if the data-set used for its training is unambiguous, and the same neural network will not be able to make good predictions if the data-set contains ambiguities. Therefore, we make sure that all ambiguities are completely eliminated from the data-set.

During simulations, we avoid all \textit{trivial} ambiguities that may be present in correlation traces \cite{Kane1993}: (\textit{i}) the direction-of-time ambiguity is intrinsically absent by the nature of cross-correlation measurements, where time symmetry is broken; (\textit{ii}) the addition-of-constant-spectral-phase ambiguity is eliminated by shifting spectral phase so that it would correspond to 0 radians at the spectral peak; (\textit{iii}) the time-shift ambiguity is avoided by making peaks of the pulse electric-field amplitudes to be located at 0-fs time and by subtracting the linear components of the spectral phases from overall phase profiles.

After having eliminated the trivial ambiguities, the only possible ambiguities left are those referred to as \textit{non-trivial}. In order to avoid non-trivial ambiguities and characterise the second pulse unambiguously (without the knowledge of the reference pulse), we influence its spectral phase in a predictable way by making it propagate through a glass plate (e.g., fused silica) of certain thickness $d$, dispersion of which is well-documented \cite{Malitson1965}. As a result of such propagation, additional phase $kd$ is added to the overall spectral phase of the pulse, with the wave vector defined as $k=n(\omega)\omega/c$, where $n(\omega)$ is the frequency-dependent refractive index of the glass, $\omega$ is the angular frequency, and $c$ is the speed of light. The dispersion $n(\omega)$ of the refractive index was estimated with the help of the Sellmeier equation for fused silica (see Supplementary material, Section S2) \cite{Malitson1965,Sellmeier1872}. The resultant electric field of the modified second pulse in the frequency domain was then Fourier-transformed to yield the electric field in the time domain affected by the glass plate (Figure~\ref{fig:data_sample}f). The corresponding cross-correlation trace obtained with this stretched pulses is shown in Figure~\ref{fig:data_sample}g

\begin{figure}[h!]
\centering \includegraphics[width=1\textwidth,height=0.8\textheight,keepaspectratio]{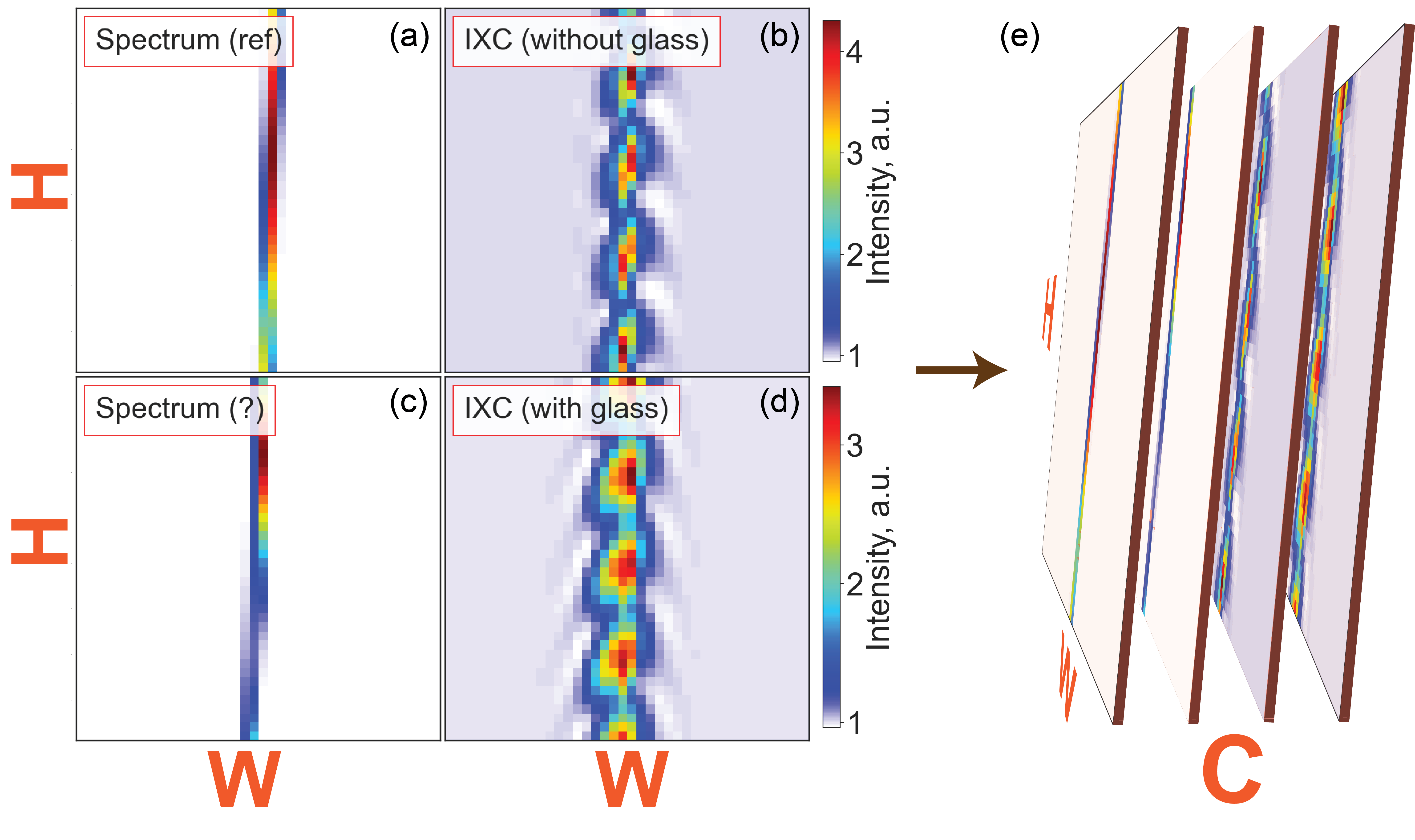}
\caption{Spectra of the reference pulse (a), and of the pulse to be characterised (c), reshaped into 2D images with dimensions $W\times H$ (see the text for details); interferometric cross-correlation traces with the 2-mm fused silica plate out (b) and in (d), reshaped into 2D images with dimensions $W\times H$. (e) Reshaped one-dimensional traces, stacked to form a three-dimensional data-sample of size $W\times H\times C$ with $W=H=40$ and $C=4$. The colormap in (a,c) is the same as in (b,d) with white color corresponding to 0 and the darkest shade of red corresponding to 1.}
\label{fig:reshaped}
\end{figure}

As clearly indicated previously \cite{Naganuma1989,Chung2001}, minimum three measurements are required to retrieve the pulses from 1D data-sets, and adding extra information can only lead to a better convergence of an iterative algorithm and a higher level of its reliability. Here, four 1D data-sets -- spectra of both pulses and two interferometric cross-correlation traces obtained with the glass plate in and out of the interferometer's arm -- were used to reconstruct electric field $E_?(t)$ of the pulse of interest. Examples of such 1D traces are shown in Figures~\ref{fig:data_sample}a,b,f,g. The choice of inclusion of spectra of both pulses as additional constraints was motivated by the dependence of the fringe contrast of interferometric cross-correlation traces on the amount of spectral overlap between the two pulses: the greater the spectral overlap between the pulses, the larger the fringe contrast in the corresponding interferometric traces. The spectral content of the pulses and the extent of their spectral overlap, therefore, should serve as a guide for the neural network to reconstruct pulses more reliably. The four 1D data-sets also imply the simplest experimental setting, in which only a single-element detector and one delay stage are used.

Finally, in order to take advantage of CNNs in solving such 1D pulse-retrieval problem, and to exploit their three-dimensional (3D) convolutional filters capable of efficient characterisation (encoding) of patterns in input data-sets, we transform the four mentioned 1D observables into 3D data-structures via two transformations referred to as \textit{reshaping} and \textit{stacking}. In the process of reshaping, all four 1D data-sets (e.g. those shown in Figures~\ref{fig:data_sample}a,b,f,g) each containing $N$ values are split into $W$ pieces of length $H$, where $N=W\times H$. Each of these pieces are then used to form columns of a matrix representing a resultant 2D image. As a result, four images of size $W\times H$  are formed (Figures~\ref{fig:reshaped}a--d). In the stacking procedure, we concatenate these four images along the third dimension forming the depth of the 3D data-structure, which we refer here to as the number of channels $C$ following convention in the machine-learning community. Structurally, a single data-sample, therefore, is a 3D data-cube of size $W\times H\times C$ with the number of channels $C$ representing the amount of one-dimensional data-sets used to retrieve pulses. Each such 3D data-sample is supplemented by the corresponding electric field of the pulse of interest (e.g. the one shown in Figure~\ref{fig:data_sample}d), which is referred to as "label" and serves as a guide for the neural network during the learning period.

In our machine-learning experiments, we used overall 20000 such data-samples. We split this data into three parts: (\textit{i}) \textit{training} data-set consisting of 19000 data-samples, (\textit{ii}) \textit{development} data-set consisting of 600 data-samples, and (\textit{iii}) \textit{test} data-set consisting of 400 data-samples. We used training data-set to train the neural network, and the accuracy of the prediction of the labels from the development data-set as a feedback parameter to improve our network's predicting ability, as will be discussed in more detail in the next section.

\section{Neural-network architecture and training}

The neural-network architecture used in this work was implemented in Python with the help of the PyTorch machine-learning framework \cite{NEURIPS2019_9015}, and, as mentioned above, is based on convolutional neural networks (CNNs) \cite{Fukushima1980,LeCun1999,Schmidhuber2015}. We adopted the simplest conceptual design of CNN with the two main parts: \textit{an encoder} and \textit{a decoder}. The encoder (consisting of convolutional layers) encodes features contained in the training data-set, and during the training stage tries to generalise trends observed in this data-set. The decoder (consisting of fully-connected layers) uses the information from the output of the encoder in order to retrieve particular information of interest (electric fields in our case). As mentioned earlier, this particular information of interest is represented as labels in the data-set. As the neural network learns, it compares its current predictions with these labels, and tries to adjust its learnt parameters by means of the (mini-batch) gradient-descent algorithm \cite{Fukushima1980,LeCun1999}  (we used the Adam optimization algorithm \cite{Kingma2014} as the corresponding gradient-descent-based optimisation routine). As a feedback for this parameter adjustment procedure, at each iteration the squared $L_2$-norm ($L$) over the development data-set is calculated as
\begin{align}
L_{dev} = \frac{1}{N} \sum_{i=1}^{N} \sum_{j=1}^{n_l} (E_{?,i,j} - \hat{E}_{?,i,j})^2,
\label{eqn:cost}
\end{align}
where $N$ is the number of data-samples in the development data-set; $n_l$ is the number of elements of a 1D array of numbers representing pulse electric field ($E_{?,i,j}$ or $\hat{E}_{?,i,j}$) in a data-sample from the development data-set; $E_{?,i,j}$ is the ground-truth electric field (label); $\hat{E}_{?,i,j}$ is the corresponding electric field predicted by the neural network at the current learning stage; $L_{dev}$ is also referred to as the objective function; fields are normalised so that $|E_{?,i,j}|\leq1$ and $|\hat{E}_{?,i,j}|\leq1$. This feedback is used to adjust parameters of the neural network such as weights and biases as well as other parameters referred to as hyperparameters. The last serve as "tuning knobs" allowing to increase the neural-network's overall performance (see Supplementary Material, Section 4, for more details on both the architecture of our neural network, the training procedure, and the hyperparameter optimisation procedure).

We trained our neural network on a computer equipped with the Titan RTX graphic processing unit (GPU) with up to 130 teraFLOPS (floating-point operations per second) and 24-gigabyte memory. Training-time rate is estimated to be $\sim$83.3~sec/epoch, while the pulse retrieval takes as short as $\sim$1.1~milliseconds.

\section{Results and Discussion}

\subsubsection{Prediction using simulated data}

\begin{figure}[h!]
\centering \includegraphics[width=1\textwidth,height=0.75\textheight,keepaspectratio]{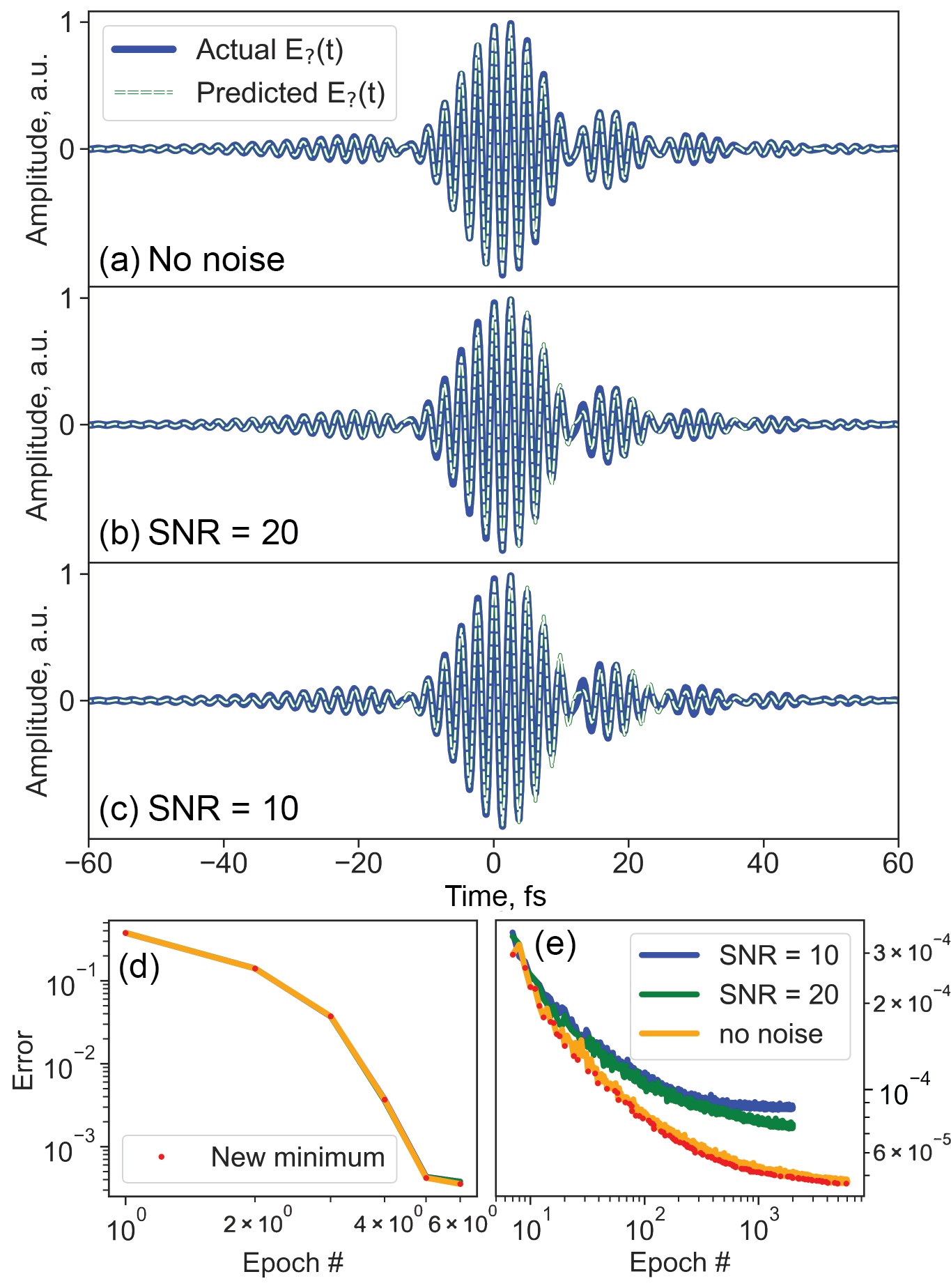}
\caption{Actual (solid line) and neural-network-predicted (dashed line) electric field in the case of the training data-set (a) without addition of noise, (b) with SNR = 20, and (c) with SNR = 10. (d,e) Convergence curves of the prediction error versus the number of epochs when the data-set was without noise (orange), with SNR = 20 (green), and with SNR = 10 (blue). Red dots mark newly-achieved minima of the error convergence curve (in the case of the training data-set without noise). The horizontal and vertical axes are given in logarithmic scale.}
\label{fig:predicted_no_noise} 
\end{figure}

Figure~\ref{fig:predicted_no_noise} illustrates the results of neural-network predictions on the simulated data. Figure~\ref{fig:predicted_no_noise}a shows an example of the neural-network-predicted electric field from the test data-set (see more examples in Supplementary material, Section S5). It is clear that the electric field envelope and phase are captured by the neural network with high accuracy, even though the shape of the pulse is not trivial, with multiple peaks in the electric field amplitude.
After 6000 epochs of training, the $L_2$-norm (Equation (\ref{eqn:cost})) over the training, development and test data-sets was achieved to be as low as $4.6\cdot10^{-5}$, $4.7\cdot10^{-5}$, and $9.2\cdot10^{-5}$, respectively. These numbers reflect the prediction error integrated over 1600 time-points and averaged over the development data-set so that the average root-mean-square (rms) error per time-point per data-sample is estimated to be of the order of 10$^{-4}$. Considering the normalization of the total pulse electric field ($|E_{?,i,j}|\leq1$) this level of rms-error is very small and underlines high prediction accuracy, which in turn signifies that there are no ambiguities present in the generated data-set. This high prediction accuracy is expected as all trivial ambiguities are eliminated by the construction of the data-set and by the nature of cross-correlation traces facilitating unambiguous mapping of the input data to the corresponding pulse electric field. If there were ambiguities in the generated data-set, it would not be possible to achieve with the current neural-network architecture by design. Figures~\ref{fig:predicted_no_noise}d,e (orange curve) show the error calculated from the predictions over the development data-set, which the neural network has not seen during training. There is a steeply-decaying learning curve during the first five epochs of training when the network learns high-amplitude part of the pulse shape, followed by a gradual decay of the error over the rest of the epochs during which low-amplitude parts (such as wings) are learned.

To test the robustness of our neural-network architecture to the presence of noise we added white Gaussian noise (see details in Supplementary material, Section S6) to the simulated data-set, and retrained the network over 2000 epochs using the noisy data-sets with two different noise levels. Figures~\ref{fig:predicted_no_noise}b,c show the predicted electric fields when the training data-set was augmented by noise yielding signal-to-noise ratio (SNR) of 20 (Figure~\ref{fig:predicted_no_noise}b) and 10 (Figure~\ref{fig:predicted_no_noise}c). In both cases, accuracy of the predictions is reasonably high, with the $L_2$-norm (Equation (\ref{eqn:cost})) of the order of $10^{-4}$ (and the average rms of the order of 10$^{-3}$). In the case of SNR = 20, the $L_2$-norm over the training, development and test data-sets was achieved to be as low as $8.9\cdot10^{-5}$, $7.4\cdot10^{-5}$, and $11.7\cdot10^{-5}$, respectively, whereas in the case of SNR = 10, the $L_2$-norm was obtained to be as low as $2.7\cdot10^{-5}$, $8.6\cdot10^{-5}$, and $7.8\cdot10^{-5}$, respectively.









\clearpage

\subsubsection{Prediction using experimental data}

\begin{figure}[h!]
\centering \includegraphics[width=1\textwidth,height=0.8\textheight,keepaspectratio]{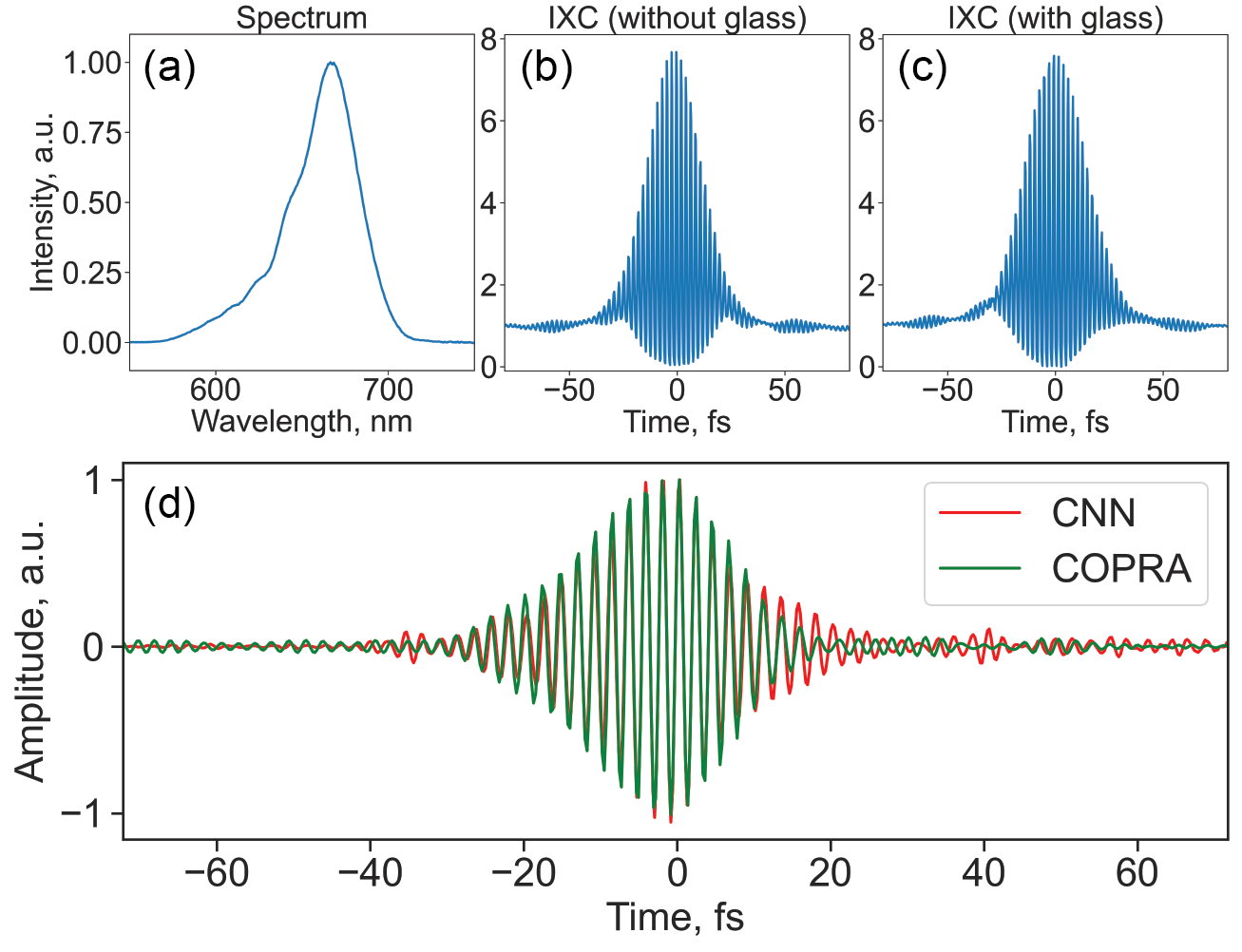}
\caption{(a--c) The experimental data used as input to the neural network: (a) spectrum of both pulses in the two arms of the interferometer; (b,c) interferometric cross-correlation trace obtained with the fused silica glass plate (b) out of and (c) in the beam (in the second arm of the interferometer). The traces in (b,c) were obtained from corresponding frequency-resolved data by integrating along the frequency axis. (d) Pulses retrieved with the neural network (red) and with COPRA \cite{Geib2019} (green).}
\label{fig:predicted_experimental}
\end{figure}

Figure~\ref{fig:predicted_experimental} summarizes the result of application of the CNN model to measured data. The data was acquired with a table-top spectrometer (Avantes) and the recently-reported fully-symmetric Michelson interferometer \cite{Kolesnichenko:20}. The measured data was used to form the neural-network input data-sample and is shown in Figures~\ref{fig:predicted_experimental}a--c. We used the same spectrum (Figure~\ref{fig:predicted_experimental}a) for the first two channels of our input experimental data-sample, as the pulses come from the same laser source (lab-built non-collinear optical parametric amplifier). The interferometric auto-correlation trace used to form the third channel of the data-sample is shown in Figure~\ref{fig:predicted_experimental}b. For interferometric cross-correlation measurements with a phase-modulated pulse to be diagnosed we break the symmetry of our interferometer by inserting a 1-mm-thick fused-silica glass plate in front of the spherical mirror in the interferometer's second arm (see Supplementary material, Section S7, for more details). The interferometric cross-correlation trace obtained in this case was used to form the fourth channel of the data-sample and is shown in Figure~\ref{fig:predicted_experimental}c. The pulse in the second arm of the interferometer passes through the fused-silica glass plate twice and should therefore be affected as if it passed a 2-mm glass plate similarly to simulation of the numerical data-set.

The 1D interferometric cross-correlation traces were derived from 2D interferometric FROG (iFROG) traces by integrating the last along the frequency axis. We acquired iFROG traces (instead of 1D correlation traces) to be able to compare the neural-network retrieval results with those obtained using recently reported COPRA pulse retrieval algorithm \cite{Geib2019} (which operates on frequency-resolved data). The iFROG traces were acquired by recording second-harmonic signal generated in a 5-$\mu$m-thick beta barium borate (BBO) crystal. The traces were recorded over the 600-fs range of time delays (between the two pulses in the two interferometer's arms), with 0.1~fs time step and 50~ms integration time. The two pulses from the output of the interferometer were focused by an off-axis parabolic mirror (with focal length of 20.3~mm) onto the BBO-crystal. When recording both the fundamental spectrum of the pulses and the second-harmonic-generation spectra during iFROG measurements, the spectrometer was placed after the position of the nonlinear crystal. The recorded experimental data was pre-processed to match the standard for inputs that the neural network accepts, as defined above.

Figure~\ref{fig:predicted_experimental}d shows the pulse field retrieved using our CNN and the COPRA \cite{Geib2019}. We derived second-harmonic-generation FROG (SHG-FROG) by filtering out (along the time axis) first and second harmonics from the measured iFROG trace, and used the SHG-FROG trace to feed the COPRA algorithm. Although the two retrieval results exhibit small mismatches, the general pulse shape is captured very well: both retrieval approaches show pulses of similar duration and carrier frequency. The apparent mismatches between the two pulse-retrieval results likely have multiple reasons. First of all, we notice that in the measured auto-correlation trace (Figure~\ref{fig:predicted_experimental}b) the background-to-peak ratio deviates from the theoretical value of 1:8 likely due to small alignment imperfections. This effect was not present in the simulated data-set and therefore may fall slightly outside of the statistical distribution of the numerical data-set used for training the neural network. 
In addition, to achieve a better match of the experimental data to statistical distribution of simulated data-set, the presence of fine structure (wiggles on the low-energy side) observed in the measured spectrum could be better taken into account by a refined data-generation procedure. Further, the insertion of the glass plate in the second arm of the interferometer relaxes otherwise-ideal 4f-imaging of the pulses back onto the grating-based beam-splitter, and therefore the interferometric cross-correlation trace may not exactly correspond to the ideal case of pulse propagation through the 2-mm glass plate. Finally, filtering the static component as well as the first and second harmonics constituting 1D correlation traces from surrounding noise may have contributed to distortions related to a particular filtering window shape. All of these possible reasons could be taken into account in the process known as the data-engineering and will be the ground for our future work. 

Nevertheless, despite all of these reasons, the demonstrated neural-network retrieval result indicates that it is possible to characterise ultrashort pulses rapidly and reliably from 1D interferometric traces. The neural network is capable of predicting variations in the pulse shape capturing such details as multiple amplitude peaks if any, and minor oscillations at the tails of the electric field envelopes, although this could be further improved by larger training data-sets and comprehensive data-engineering taking into account the sources of experimental imperfections. Our approach is also scalable: various glasses (e.g. BK7) of various thicknesses can be used as well as other frequency ranges and pulse durations, which all could be considered within the statistical distribution of the training data-samples.








\section{Conclusions}

In conclusions, here we demonstrate the proof-of-principle approach to ultrashort-optical-pulse characterisation from 1D interferometric cross-correlation time traces using a deep neural network. We show that it is enough to measure interferometric cross-correlation between the two pulses, interferometric cross-correlation between the pulses with one of them propagating through a dispersive element, and spectra of both pulses. Generally, no knowledge about the reference pulse other than its spectrum is needed. All of this implies a simple experimental setup with only one delay stage and a single-channel detector, which together with milliseconds-long pulse retrieval makes ultrashort pulse characterisation a fast procedure. This could be further improved up to real-time  pulse characterisation technique by smart design of the experimental setup where both cross-correlation traces are acquired together and in a real-time acquisition using sweeping-mode of the delay stages \cite{Fork:78}.





\section*{Acknowledgements}
We thank Nils C. Geib for the support provided for using the PyPret-package \cite{pypret} containing the COPRA algorithm \cite{Geib2019} for pulse retrieval. The work was supported by Vetenskapsrådet, Crafoordska Stiftelsen, and NanoLund.

\section*{Disclosures}
The authors declare no conflicts of interest.

\clearpage
\printbibliography

@Article{Yan1991,
  author    = {Chi Yan and Jean-Claude Diels},
  journal   = {Journal of the Optical Society of America B},
  title     = {Amplitude and phase recording of ultrashort pulses},
  year      = {1991},
  month     = {jun},
  number    = {6},
  pages     = {1259--1263},
  volume    = {8},
  doi       = {10.1364/josab.8.001259},
  publisher = {The Optical Society},
}

@Article{Chung2001,
  author    = {Jung-Ho Chung and Andrew M. Weiner},
  journal   = {{IEEE} Journal of Selected Topics in Quantum Electronics},
  title     = {Ambiguity of ultrashort pulse shapes retrieved from the intensity autocorrelation and the power spectrum},
  year      = {2001},
  number    = {4},
  pages     = {656--666},
  volume    = {7},
  doi       = {10.1109/2944.974237},
  publisher = {Institute of Electrical and Electronics Engineers ({IEEE})},
}

@Article{Diels1985,
  author    = {Jean-Claude M. Diels and Joel J. Fontaine and Ian C. McMichael and Francesco Simoni},
  journal   = {Applied Optics},
  title     = {Control and measurement of ultrashort pulse shapes (in amplitude and phase) with femtosecond accuracy},
  year      = {1985},
  month     = {may},
  number    = {9},
  pages     = {1270--1282},
  volume    = {24},
  doi       = {10.1364/ao.24.001270},
  publisher = {The Optical Society},
}

@InProceedings{Diels1985a,
  author    = {Jean-Claude Diels},
  booktitle = {Proceedings of SPIE},
  title     = {Measurement techniques with mode-locked dye laser},
  year      = {1985},
  editor    = {M. J. Soileau},
  month     = {apr},
  pages     = {63--70},
  publisher = {{SPIE}},
  volume    = {533},
  doi       = {10.1117/12.946542},
}

@InProceedings{DIELS:87,
  author    = {J.-C. M. Diels and J. J. Fontaine and N. Jamasbi and Ming Lai and J. Mackey},
  booktitle = {Conference on Lasers and Electro-Optics},
  title     = {Femto-nitpicker},
  year      = {1987},
  pages     = {MD3},
  publisher = {Optical Society of America},
  journal   = {Conference on Lasers and Electro-Optics},
  url       = {http://www.osapublishing.org/abstract.cfm?URI=CLEO-1987-MD3},
}

@Article{Gerchberg72,
  author              = {Gerchberg, R. W. and Saxton, W. O.},
  journal             = {OPTIK},
  title               = {A practical algorithm for the determination of phase from image and diffraction plane pictures},
  year                = {1972},
  issn                = {{0030-4026}},
  number              = {2},
  pages               = {237--246},
  volume              = {35},
  doc-delivery-number = {{M3124}},
  journal-iso         = {{Optik}},
  type                = {{Article}},
  unique-id           = {{ISI:A1972M312400012}},
}

@Article{Peatross1998,
  author    = {J. Peatross and A. Rundquist},
  journal   = {Journal of the Optical Society of America B},
  title     = {Temporal decorrelation of short laser pulses},
  year      = {1998},
  month     = {jan},
  number    = {1},
  pages     = {216--222},
  volume    = {15},
  doi       = {10.1364/josab.15.000216},
  publisher = {The Optical Society},
}

@Article{Kakarala1993,
  author    = {Ramakrishna Kakarala and Geoffrey J. Iverson},
  journal   = {Journal of the Optical Society of America A},
  title     = {Uniqueness of results for multiple correlations of periodic functions},
  year      = {1993},
  month     = {jul},
  number    = {7},
  pages     = {1517--1528},
  volume    = {10},
  doi       = {10.1364/josaa.10.001517},
  publisher = {The Optical Society},
}

@Article{Naganuma1990,
  author    = {Kazunori Naganuma and Kazuo Mogi and Hajime Yamada},
  journal   = {Optics Letters},
  title     = {Group-delay measurement using the Fourier transform of an interferometric cross correlation generated by white light},
  year      = {1990},
  month     = {apr},
  number    = {7},
  pages     = {393--395},
  volume    = {15},
  doi       = {10.1364/ol.15.000393},
  publisher = {The Optical Society},
}

@Article{Naganuma1989,
  author    = {K. Naganuma and K. Mogi and H. Yamada},
  journal   = {{IEEE} Journal of Quantum Electronics},
  title     = {General method for ultrashort light pulse chirp measurement},
  year      = {1989},
  month     = {jun},
  number    = {6},
  pages     = {1225--1233},
  volume    = {25},
  doi       = {10.1109/3.29252},
  publisher = {Institute of Electrical and Electronics Engineers ({IEEE})},
}

@Article{Naganuma1989a,
  author    = {Kazunori Naganuma and Kazuo Mogi and Hajime Yamada},
  journal   = {Applied Physics Letters},
  title     = {Time direction determination of asymmetric ultrashort optical pulses from second-harmonic generation autocorrelation signals},
  year      = {1989},
  month     = {mar},
  number    = {13},
  pages     = {1201--1202},
  volume    = {54},
  doi       = {10.1063/1.100752},
  publisher = {{AIP} Publishing},
}

@Article{Zahavy2018,
  author    = {Tom Zahavy and Alex Dikopoltsev and Daniel Moss and Gil Ilan Haham and Oren Cohen and Shie Mannor and Mordechai Segev},
  journal   = {Optica},
  title     = {Deep learning reconstruction of ultrashort pulses},
  year      = {2018},
  month     = {may},
  number    = {5},
  pages     = {666--673},
  volume    = {5},
  doi       = {10.1364/optica.5.000666},
  publisher = {The Optical Society},
}

@Article{Kleinert2019,
  author    = {Sven Kleinert and Ayhan Tajalli and Tamas Nagy and Uwe Morgner},
  journal   = {Optics Letters},
  title     = {Rapid phase retrieval of ultrashort pulses from dispersion scan traces using deep neural networks},
  year      = {2019},
  month     = {feb},
  number    = {4},
  pages     = {979--982},
  volume    = {44},
  doi       = {10.1364/ol.44.000979},
  publisher = {The Optical Society},
}

@Article{Krumbuegel1996,
  author    = {Marco A. Krumbügel and Celso L. Ladera and Kenneth W. DeLong and David N. Fittinghoff and John N. Sweetser and Rick Trebino},
  journal   = {Optics Letters},
  title     = {Direct ultrashort-pulse intensity and phase retrieval by frequency-resolved optical gating and a computational neural network},
  year      = {1996},
  month     = {jan},
  number    = {2},
  pages     = {143--145},
  volume    = {21},
  doi       = {10.1364/ol.21.000143},
  publisher = {The Optical Society},
}

@Article{Fukushima1980,
  author    = {Kunihiko Fukushima},
  journal   = {Biological Cybernetics},
  title     = {Neocognitron: A self-organizing neural network model for a mechanism of pattern recognition unaffected by shift in position},
  year      = {1980},
  month     = {apr},
  number    = {4},
  pages     = {193--202},
  volume    = {36},
  doi       = {10.1007/bf00344251},
  publisher = {Springer Science and Business Media {LLC}},
}

@InCollection{LeCun1999,
  author    = {Yann LeCun and Patrick Haffner and L{\'{e}}on Bottou and Yoshua Bengio},
  booktitle = {Shape, Contour and Grouping in Computer Vision},
  publisher = {Springer Berlin Heidelberg},
  title     = {Object recognition with gradient-based learning},
  year      = {1999},
  pages     = {319--345},
  doi       = {10.1007/3-540-46805-6_19},
}

@Article{Schmidhuber2015,
  author    = {Jürgen Schmidhuber},
  journal   = {Neural Networks},
  title     = {Deep learning in neural networks: An overview},
  year      = {2015},
  month     = {jan},
  pages     = {85--117},
  volume    = {61},
  doi       = {10.1016/j.neunet.2014.09.003},
  publisher = {Elsevier {BV}},
}

@InProceedings{Noble2020,
  author    = {Joshua Noble and Chen Zhou and William T. Murray and Zhiwen Liu},
  booktitle = {Ultrafast Nonlinear Imaging and Spectroscopy {VIII}},
  title     = {Convolutional neural network reconstruction of ultrashort optical pulses},
  year      = {2020},
  editor    = {Zhiwen Liu and Demetri Psaltis and Kebin Shi},
  month     = {aug},
  pages     = {11497},
  publisher = {Proc. SPIE},
  series    = {114970L},
  doi       = {10.1117/12.2571172},
}

@Article{Ziv2020,
  author    = {Ron Ziv and Alex Dikopoltsev and Tom Zahavy and Ittai Rubinstein and Pavel Sidorenko and Oren Cohen and Mordechai Segev},
  journal   = {Optics Express},
  title     = {Deep learning reconstruction of ultrashort pulses from 2D spatial intensity patterns recorded by an all-in-line system in a single-shot},
  year      = {2020},
  month     = {feb},
  number    = {5},
  pages     = {7528--7538},
  volume    = {28},
  doi       = {10.1364/oe.383217},
  publisher = {The Optical Society},
}

@Article{Xiong2020,
  author    = {Wen Xiong and Brandon Redding and Shai Gertler and Yaron Bromberg and Hemant D. Tagare and Hui Cao},
  journal   = {{APL} Photonics},
  title     = {Deep learning of ultrafast pulses with a multimode fiber},
  year      = {2020},
  month     = {sep},
  number    = {9},
  pages     = {096106},
  volume    = {5},
  doi       = {10.1063/5.0007037},
  publisher = {{AIP} Publishing},
}

@Article{Zhu2020,
  author    = {Zheyuan Zhu and Jonathon White and Zenghu Chang and Shuo Pang},
  journal   = {Scientific Reports},
  title     = {Attosecond pulse retrieval from noisy streaking traces with conditional variational generative network},
  year      = {2020},
  month     = {apr},
  number    = {1},
  volume    = {10},
  doi       = {10.1038/s41598-020-62291-6},
  publisher = {Springer Science and Business Media {LLC}},
}

@Article{Kane1993,
  author    = {D.J. Kane and R. Trebino},
  journal   = {{IEEE} Journal of Quantum Electronics},
  title     = {Characterization of arbitrary femtosecond pulses using frequency-resolved optical gating},
  year      = {1993},
  number    = {2},
  pages     = {571--579},
  volume    = {29},
  doi       = {10.1109/3.199311},
  publisher = {Institute of Electrical and Electronics Engineers ({IEEE})},
}

@Article{Miranda2011,
  author    = {Miguel Miranda and Thomas Fordell and Cord Arnold and Anne L'Huillier and Helder Crespo},
  journal   = {Optics Express},
  title     = {Simultaneous compression and characterization of ultrashort laser pulses using chirped mirrors and glass wedges},
  year      = {2011},
  month     = {dec},
  number    = {1},
  pages     = {688--697},
  volume    = {20},
  doi       = {10.1364/oe.20.000688},
  publisher = {The Optical Society},
}

@Article{Iaconis1998,
  author    = {C. Iaconis and I. A. Walmsley},
  journal   = {Optics Letters},
  title     = {Spectral phase interferometry for direct electric-field reconstruction of ultrashort optical pulses},
  year      = {1998},
  month     = {may},
  number    = {10},
  pages     = {792--794},
  volume    = {23},
  doi       = {10.1364/ol.23.000792},
  publisher = {The Optical Society},
}

@Article{Lozovoy2004,
  author    = {Vadim V. Lozovoy and Igor Pastirk and Marcos Dantus},
  journal   = {Optics Letters},
  title     = {Multiphoton intrapulse interference{\hspace{1em}}{IV}{\hspace{1em}}Ultrashort laser pulse spectral phase characterization and compensation},
  year      = {2004},
  month     = {apr},
  number    = {7},
  pages     = {775--777},
  volume    = {29},
  doi       = {10.1364/ol.29.000775},
  publisher = {The Optical Society},
}

@Book{Trebino2000,
  author    = {Rick Trebino},
  publisher = {Springer {US}},
  title     = {Frequency-Resolved Optical Gating: The Measurement of Ultrashort Laser Pulses},
  year      = {2000},
  doi       = {10.1007/978-1-4615-1181-6},
}

@Article{Geib2019,
  author    = {Nils C. Geib and Matthias Zilk and Thomas Pertsch and Falk Eilenberger},
  journal   = {Optica},
  title     = {Common pulse retrieval algorithm: a fast and universal method to retrieve ultrashort pulses},
  year      = {2019},
  month     = {apr},
  number    = {4},
  pages     = {495--505},
  volume    = {6},
  doi       = {10.1364/optica.6.000495},
  publisher = {The Optical Society},
}

@InProceedings{10.5555/2998687.2998822,
  author    = {Smyth, Padhraic and Fayyad, Usama and Burl, Michael and Perona, Pietro and Baldi, Pierre},
  booktitle = {Proceedings of the 7th International Conference on Neural Information Processing Systems},
  title     = {Inferring Ground Truth from Subjective Labelling of Venus Images},
  year      = {1994},
  address   = {Cambridge, MA, USA},
  pages     = {1085–1092},
  publisher = {MIT Press},
  series    = {NIPS'94},
  location  = {Denver, Colorado},
  numpages  = {8},
}

@Article{Gao2017,
  author    = {Bin-Bin Gao and Chao Xing and Chen-Wei Xie and Jianxin Wu and Xin Geng},
  journal   = {{IEEE} Transactions on Image Processing},
  title     = {Deep Label Distribution Learning With Label Ambiguity},
  year      = {2017},
  month     = {jun},
  number    = {6},
  pages     = {2825--2838},
  volume    = {26},
  doi       = {10.1109/tip.2017.2689998},
  publisher = {Institute of Electrical and Electronics Engineers ({IEEE})},
}

@incollection{NEURIPS2019_9015,
title = {PyTorch: An Imperative Style, High-Performance Deep Learning Library},
author = {Paszke, Adam and Gross, Sam and Massa, Francisco and Lerer, Adam and Bradbury, James and Chanan, Gregory and Killeen, Trevor and Lin, Zeming and Gimelshein, Natalia and Antiga, Luca and Desmaison, Alban and Kopf, Andreas and Yang, Edward and DeVito, Zachary and Raison, Martin and Tejani, Alykhan and Chilamkurthy, Sasank and Steiner, Benoit and Fang, Lu and Bai, Junjie and Chintala, Soumith},
booktitle = {Advances in Neural Information Processing Systems 32},
editor = {H. Wallach and H. Larochelle and A. Beygelzimer and F. d\textquotesingle Alch\'{e}-Buc and E. Fox and R. Garnett},
pages = {8024--8035},
year = {2019},
publisher = {Curran Associates, Inc.},
url = {http://papers.neurips.cc/paper/9015-pytorch-an-imperative-style-high-performance-deep-learning-library.pdf}
}

@Article{Sellmeier1872,
  author    = {W. Sellmeier},
  journal   = {Annalen der Physik und Chemie},
  title     = {Ueber die durch die Aetherschwingungen erregten Mitschwingungen der Körpertheilchen und deren Rückwirkung auf die ersteren, besonders zur Erklärung der Dispersion und ihrer Anomalien},
  year      = {1872},
  number    = {11},
  pages     = {386--403},
  volume    = {223},
  doi       = {10.1002/andp.18722231105},
  publisher = {Wiley},
}

@Article{Malitson1965,
  author    = {I. H. Malitson},
  journal   = {Journal of the Optical Society of America},
  title     = {Interspecimen Comparison of the Refractive Index of Fused Silica},
  year      = {1965},
  month     = {oct},
  number    = {10},
  pages     = {1205},
  volume    = {55},
  doi       = {10.1364/josa.55.001205},
  publisher = {The Optical Society},
}

@Article{Kingma2014,
  author        = {Diederik P. Kingma and Jimmy Ba},
  title         = {Adam: A Method for Stochastic Optimization},
  year          = {2014},
  month         = dec,
  archiveprefix = {arXiv},
  eprint        = {1412.6980},
  primaryclass  = {cs.LG},
}

@Misc{pypret,
  author = {Nils C. Geib},
  title  = {Python for pulse retrieval},
  url    = {https://github.com/ncgeib/pypret},
}

@article{Fork:78,
author = {R. L. Fork and F. A. Beisser},
journal = {Appl. Opt.},
number = {22},
pages = {3534--3535},
publisher = {OSA},
title = {Real-time intensity autocorrelation interferometer},
volume = {17},
month = {Nov},
year = {1978},
url = {http://www.osapublishing.org/ao/abstract.cfm?URI=ao-17-22-3534},
doi = {10.1364/AO.17.003534},
}

@article{Kolesnichenko:20,
author = {Pavel V. Kolesnichenko and Lukas Wittenbecher and Donatas Zigmantas},
journal = {Opt. Express},
number = {25},
pages = {37752--37757},
publisher = {OSA},
title = {Fully symmetric dispersionless stable transmission-grating Michelson interferometer},
volume = {28},
month = {Dec},
year = {2020},
url = {http://www.opticsexpress.org/abstract.cfm?URI=oe-28-25-37752},
doi = {10.1364/OE.409185},
}

\end{document}